\documentclass[%
 aip,
 amsmath,amssymb,
 reprint,%
]{revtex4-1}

\usepackage{graphicx}
\usepackage{dcolumn}
\usepackage{bm}
\usepackage{xcolor}
\usepackage[utf8]{inputenc}
\usepackage[T1]{fontenc}
\usepackage{mathptmx}
\usepackage{etoolbox}

\makeatletter
\def\@email#1#2{%
 \endgroup
 \patchcmd{\titleblock@produce}
  {\frontmatter@RRAPformat}
  {\frontmatter@RRAPformat{\produce@RRAP{*#1\href{mailto:#2}{#2}}}\frontmatter@RRAPformat}
  {}{}
}%
\makeatother
\begin{document}

\preprint{AIP/123-QED}

\title{Simultaneous measurement of specific heat and thermal conductivity in
pulsed magnetic fields}%

\author{Tetsuya Nomoto}%
\affiliation{ 
Institute for Solid State Physics, The University of Tokyo, Kashiwa, Chiba 277-8581, Japan
}%
\author{Chengchao Zhong}%
\author{Hiroshi Kageyama}%
\affiliation{ 
Department of Energy and Hydrocarbon Chemistry, Graduate School of Engineering, Kyoto University, Kyoto 615-8510, Japan
}%
\author{Yoko Suzuki}%
\affiliation{ 
MPA MAGLAB, Los Alamos National Laboratoy, Los Alamos, NM 87545, U. S. A.
}%
\author{Marcelo Jaime}%
\affiliation{ 
MPA MAGLAB, Los Alamos National Laboratoy, Los Alamos, NM 87545, U. S. A.
}%
\author{Yoshiaki Hashimoto}%
\author{Shingo Katsumoto}%
\author{Naofumi Matsuyama}%
\affiliation{ 
Institute for Solid State Physics, The University of Tokyo, Kashiwa, Chiba 277-8581, Japan
}%
\author{Chao Dong}%
\altaffiliation{
Present address: Wuhan National High Magnetic Field Center, Huazhong University of Science and Technology, Wuhan 430074, China
}
\affiliation{ 
Institute for Solid State Physics, The University of Tokyo, Kashiwa, Chiba 277-8581, Japan
}%
\author{Akira Matsuo}%
\author{Koichi Kindo}%
\affiliation{ 
Institute for Solid State Physics, The University of Tokyo, Kashiwa, Chiba 277-8581, Japan
}%
\author{Koichi Izawa}%
\affiliation{ 
Division of Material Physics, Graduate School of Engineering Science, Osaka University, Toyonaka, Osaka 560-8531, Japan
}%
\author{Yoshimitsu Kohama}%
\email{ykohama@issp.u-tokyo.ac.jp}
\affiliation{ 
Institute for Solid State Physics, The University of Tokyo, Kashiwa, Chiba 277-8581, Japan
}%

\date{\today}

\begin{abstract}
We report an experimental setup for simultaneously measuring specific heat and thermal conductivity in feedback-controlled pulsed magnetic fields of 50 msec duration at cryogenic temperatures. A stabilized magnetic field pulse obtained by the feedback control, which dramatically improves the thermal stability of the setup and sample, is used in combination with the \textit{flash} method to obtain absolute values of thermal properties up to ~37.2 T in the 2 K to 16 K temperature range. We describe the  experimental setup and demonstrate the performance of the present method with measurements on single crystal samples of the geometrically frustrated quantum spin-dimer system SrCu$_2$(BO$_3$)$_2$. Our proof-of-principle results show excellent agreement with data taken using a standard steady-state method, confirming the validity and convenience of the present approach.

\end{abstract}

\maketitle
\section{\label{sec:level1}Introduction}

Thermal properties such as specific heat and thermal conductivity are of fundamental importance in understanding physical phenomena of matter,\cite{R1, R2} as well as in applications utilizing thermal devices.\cite{R3,R4} While specific heat is a sensitive probe both for itinerant and localized excitations, thermal conductivity is a selective probe for itinerant-type excitations and their scattering mechanisms. A comparison between specific heat and thermal conductivity can be used to identify the type of thermal excitations and provides a clear understanding of how the excited quasi-particles contribute to the physical properties. Thus, the simultaneous measurement of specific heat and thermal conductivity is a highly demanded capability in the toolbox to study physical phenomena like charge-neutral fermions,\cite{R5} unconventional superconductivity,\cite{R6,R7} and the elusive quantum spin liquid state,\cite{R8,R9,R10,RR1} among many others. However, most of the traditional measurement techniques used at low temperatures and magnetic fields can only provide one of these thermal properties. Although  specific heat and thermal conductivity measurements can be performed separately, in practice it is hard to replicate exact values for the control parameters, such as cooling/heating rate, sample orientation in the magnetic field, and quality of the samples, leading to some ambiguity in the interpretation of results. The simultaneous measurement of thermal conductivity and specific heat is then a highly desired experimental approach.

Among many measurement techniques in the literature, the technique called \textit{flash} method (or laser flash method) is commonly used to obtain thermal properties at about room temperature.\cite{R11} This technique can provide absolute values of thermal conductivity ($\kappa$), and volumetric specific heat ($c_V$) via the measurement of thermal diffusivity, $\alpha = \kappa$/$c_V$. In a traditional setup for the flash method, the heat is applied with a pulsed laser on one side of the sample and the resultant transient temperature change is monitored with an optical method on the opposite side of the sample. Monitoring temperature with an optical technique is however rather challenging at cryogenic temperatures. Thus, the modified version of the flash method was developed to investigate thermal transport properties at very low temperatures down to $\sim$0.5 K.\cite{R12,R13} This method was used to investigate the diffusive heat propagation phenomenon as well as the ballistic heat transport, where the measurement of ballistic phonon was performed within a timescale of a few microseconds. The measurement of quantum oscillation was also achieved in the past by a similar version of the flash method.\cite{R14} The wide applicability and the short measurement timescale of this method motivated us to apply it for the simultaneous measurements of specific heat and thermal conductivity under pulsed high magnetic fields.

The measurement of $c_V(T)$ and $\kappa(T)$ in pulsed magnetic fields is a challenging task due to the millisecond timescale of pulsed magnetic fields and the need for magnetic field stabilization to avoid undesired sample temperature changes due to Eddy-current heating, vortex motion in superconductors,  and/or the magnetocaloric effect. Earlier work at the National High Magnetic Fields Laboratory \cite{R15,R16} demonstrated $c_V(T)$ measurements up to 60 T, carried out in a stabilized magnetic field plateau that was $\sim$100 ms long. This work was followed by various important technique advances \cite{R17,R18} and the adoption of different measurement approaches with the feedback-controlled field stabilization technique for widening the measurable conditions. \cite{R19,R20,R26,R27,AR2} On the other hand, measurements of thermal conductivity in pulsed magnetic fields have yet to be demonstrated. With the recent improvement of the field stabilization technique that fully removes the risk of the temperature instability induced by the coupling between magnetic fields and sample temperature,\cite{R21, R22} it likely becomes possible to measure thermal conductivity under pulsed magnetic fields.

In this paper, we demonstrate that the flash method can be operated in highly stabilized pulsed magnetic fields for the simultaneous measurements of specific heat and thermal conductivity. To perform the experiment within the short timescale of the stabilized field region, we used thin-film thermometers and heaters, similar to a ballistic phonon experiment in a steady field. This enables us to measure the dynamic response of the sample temperature with a timescale of a few to hundred milliseconds, where the minimum and maximum measurable timescales are limited by the electronic devices and the duration of the stabilized magnetic fields used in this research. The heat pulse data obtained from the test sample, SrCu$_2$(BO$_3$)$_2$, were analyzed based on the half-time method and the curve fitting method,\cite{R23,R24} which are compared with the reported data taken by the steady-state method. The obtained specific heat and thermal conductivity show good agreement with the data reported in previous studies, which proves the validity of the present method for the investigations of thermal properties in high magnetic fields.

\section{\label{sec:level1}Measurement setup}
\subsection{\label{sec:level2}Construction of the probe}

A simplified schematic of our probe is shown in Fig. 1. A bar-shaped single crystal of SrCu$_2$(BO$_3$)$_2$ grown by the traveling solvent floating zone method \cite{AR1} was used as a test sample for this research. Radio frequency sputtering was used to deposit a Au-Ge film thermometer on the sample plane perpendicular to its long axis, and a Ni-Cr film heater was sputtered on the opposite end, respectively. In this configuration, heat flows along the long axis of the bar-shaped sample as indicated by the red arrow in the inset of Fig. 1. The sample was then mounted on a second piece of SrCu$_2$(BO$_3$)$_2$ which was used as a thermal bath. We note that an insulating substrate such as quartz can be used for a thermal bath when the sample does not show a large magnetocaloric effect. In this work, the thermal bath made of SrCu$_2$(BO$_3$)$_2$ minimizes the temperature difference between the sample and thermal bath induced by the large magnetocaloric effect in this material.\cite{R25} In our experiment setup, heat flows parallel to the $a$-axis of the crystal, and magnetic fields were applied parallel to the $c$-axis. To confirm the validity of our setup, we measured two different SrCu$_2$(BO$_3$)$_2$ single crystals with different setups, setup 1 and 2. The length of the long axis, $L$, and the sample weight, $W$, of the bar-shaped samples were $L$ = 1.44 mm and $W$ = 602 {\textmu}g (setup 1) and $L$ = 1.25 mm and $W$ = 583 {\textmu}g (setup 2). The Au-Ge thermometers were calibrated using the RuO$_2$ reference thermometer which was placed near the sample. To get the correct temperature of the sample and its time dependence under magnetic fields, the magnetoresistance of the Au-Ge thermometer was calibrated by measuring the isothermal magnetoresistance of the thermometer in helium exchange gas or liquid helium using pulsed magnetic fields with a pulse duration of $\sim$1.2 s. The pick-up coil, which was used to measure an induced voltage proportional to the time derivative of magnetic fields (d$B$/d$t$), was placed near the sample. The sample was cooled by liquid $^4$He stored in the bottom of G10 tube via a weak thermal link of electrical wires made of 100 {\textmu}m diameter copper and 30 {\textmu}m constantan. Except during the calibration process, the sample space was evacuated to maintain the quasi-adiabatic condition.

\begin{figure}
\includegraphics{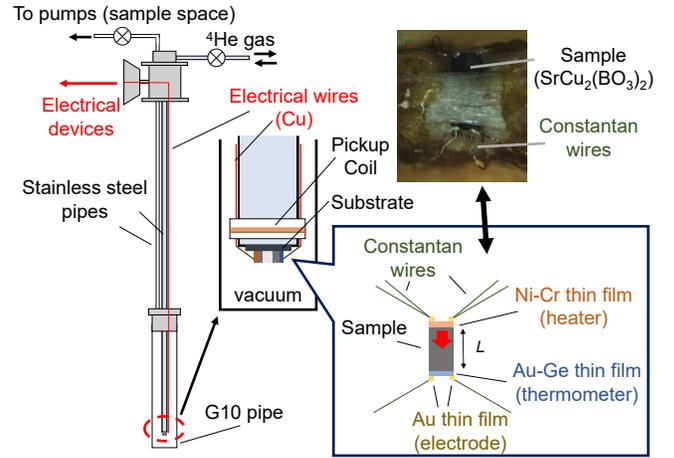}
\caption{\label{fig:epsart1} Schematic illustration of the measurement probe. The bar-shaped sample is on the bottom end of the probe.}
\end{figure}

\subsection{\label{sec:level2}Setup of electronics}

A block diagram of the electronics used in this research is shown in Fig. 2. A PXIe-6368 multifunction data acquisition board (National Instruments) was used both as a digitizer and as a generator of excitation voltage. The PXIe-6368 board can collect low-voltage signals at a 2 MHz sampling rate, which is  fast enough for the thermodynamic measurements under pulsed magnetic fields of short duration. SR560 preamplifiers (Stanford Research) were used to amplify and filter the voltage signals on the thermometer and heater. The film thermometer and film heater on the sample were connected by a quasi-four-terminal arrangement. A rectangular profile voltage, $V_{\rm{ex1}}$, in the heater circuit was generated to apply a heat pulse to the sample. The voltage, $V_{\rm{h1}}$, between the two leads of the quasi-four-terminals and the current, $I_{\rm{h}}$, through the shunt resistor, $R_{\rm{s1}}$ (= 100, 500, or 1000 $\Omega$), was observed. The $I_{\rm{h}}$ was calculated by dividing the voltage across the shunt resistor, $V_{\rm{h2}}$, by $R_{\rm{s1}}$ ($I_{\rm{h}} = V_{\rm{h2}}/ R_{\rm{s1}}$). The total applied energy, ${\Delta}Q$, was calculated by ${\Delta}Q = I_{\rm{h}} \times V_{\rm{h1}} \times \tau_{\rm{h}}$ where $\tau_{\rm{h}}$ is the duration time of pulse heating. The temperature signals, \em{i.e.}\rm, amplitudes and phases of the sinusoidal voltage between two terminals of thermometer, $V_{\rm{h1}}$($t$)$= V_{\rm{0}} \times$ sin(${\omega}t$), and the current flow through the shunt resistor ($R_{\rm{s2}}$), $I_{\rm{t}}$($t$)$  = I_{\rm{0}} \times$ sin(${\omega}t$) were determined by a digital Lock-In. Here, $I_{\rm{t}}$($t$) was obtained from the voltage across the shunt resistor, $V_{\rm{t2}}$. Since the phase shift was negligibly small at the measurement frequency of 10 kHz, the thermometer resistance, $R_{\rm{t}}$, was calculated by the ratio of those amplitudes ($R_{\rm{t}} = V_{\rm{0}}/I_{\rm{0}}$). From $R_{\rm{t}}$($t$), we obtained in this way the time dependence of sample temperature in magnetic fields.

\begin{figure}
\includegraphics{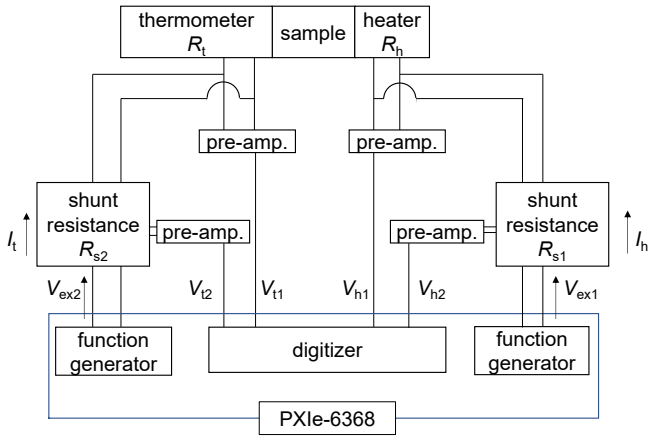}
\caption{\label{fig:epsart1} Block diagram for the electronics used in the flash method under pulsed magnetic fields.}
\end{figure}

\subsection{\label{sec:level2}Generation of flat-top magnetic fields}

To stabilize measurement temperature and improve its accuracy, we generated a so-called flat-top pulse where the magnetic field was stabilized within 0.005 T for several tens to a few hundred milliseconds. The field stabilization was performed by a proportional-integral-derivative (PID) feedback controller.\cite{R19} The feedback-controlled flat-top pulse has enabled us to measure a variety of physical quantities under high magnetic fields such as specific heat,\cite{R20,R21,R22} nuclear magnetic resonance (NMR),\cite{R26,AR2} current-voltage characteristic,\cite{R27} etc. The field feedback system consists of the main magnet and the auxiliary homemade mini magnet which compensates for the time dependence of the magnetic field originating from the main magnet. The main magnet used for this research generates a magnetic field of up to $\sim$40 T for 1.2 s, which is driven by 51.3 MW DC flywheel generator installed at the International MegaGauss Science Laboratory, Institute for Solid State Physics, The University of Tokyo. The mini magnet is energized by four or six 12 V lead-acid batteries connected in series. The current flow in the mini magnet is controlled by an insulated gate bipolar transistor (IGBT, 1MB3600VD-170E-50, Fuji Electric) and a field-programmable gate array (FPGA) module (National Instruments USB-7856R). We note that control experiments in zero magnetic field were performed in a physical properties measurement system (PPMS) by Quantum Design.

\section{\label{sec:level1}Method used for data analysis}
\subsection{\label{sec:level2}Half-time method}

The half-time method is one of the sophisticated approaches to extracting the thermal conductivity of solids from the dynamical and diffusive heat transfer processes.\cite{R28, R11} Let us now consider a transient heat transfer phenomenon along a rectangular bar of the length $L$ (see Fig. 3(a)). When a heat pulse is applied to one end of the rectangular bar ($x$ = 0) at $t$ = 0, the time dependence of the temperature at the opposite end ($x = L$), $T$($t$, $L$), is given as\cite{R28}:

\begin{equation}
T\left(t,L\right)=T_0+\ T_{max}\left[1+2\sum_{n=1}^{\infty}\left(-1\right)^n\exp{\left(-n^2\frac{t}{t_0}\right)}\right],
\end{equation}

\begin{equation}
t_0=\frac{L^2}{\alpha\pi^2},
\end{equation}

where $\alpha$ is the thermal diffusivity, $T_{\rm{0}}$ is the base temperature, and $T_{\rm{max}}$ is the maximum temperature rise. In actual measurements, $T_{\rm{0}}$ is determined by the initial temperature at $t$ = 0. The temperature change, ${\Delta}T$ = $T$($t$, $L$)-$T_{\rm{0}}$, normalized by $T_{\rm{max}}$ is plotted in Fig. 3(b) as a function of $t$/$t_{\rm{0}}$. In the adiabatic condition, $T_{\rm{max}}$ can be described as:

\begin{equation}
T_{\rm{max}}=\frac{{\Delta}Q}{C},
\end{equation}

where ${\Delta}Q$ is the applied heat and $C$ is the heat capacity of the whole rectangular bar. $C$ can be described by using the volumetric specific heat, $c_V$, as below:

\begin{equation}
C=\frac{Wc_V}{\rho},
\end{equation}

where $W$ is the sample weight and $\rho$ is the density of the sample. 

The thermal diffusivity and thermal conductivity can be simultaneously obtained by analyzing the process of temperature increase shown in Fig.3(b). The half time, $t_{\rm{1/2}}$, is the time required for ${\Delta}T$ to increase the temperature up to one half of $T_{\rm{max}}$, i.e., ${\Delta}T$ /$T_{\rm{max}}$ = 0.5, which is calculated as:

\begin{equation}
t_{1/2}=\frac{1.370L^2}{\alpha\pi^2}.
\end{equation}

Therefore, the thermal diffusivity is written as,

\begin{equation}
\alpha=\frac{0.1388L^2}{t_{1/2}}.
\end{equation}

Using the thermal diffusivity and the volumetric specific heat, the thermal conductivity, $\kappa$, is obtained as the following equation:

\begin{equation}
\kappa={\alpha}c_V=\frac{{\alpha}C{\rho}}{W}.
\end{equation}

Using the Eqs. (5), (6), and (7), $\alpha$ and $\kappa$ can be determined with  a separate estimation of $C$ by Eq. (3). In the above discussion, an adiabatic condition where no heat exchange between sample and surroundings is assumed. When the heat exchange is not negligible, ${\Delta}T$ decays proportionally to exp(- $t$/$\tau$) after showing a temperature peak. Here, $\tau$ is the relaxation time constant for reaching thermal equilibrium between the sample and surroundings. The thermal relaxation leads to a systematic error in the thermal conductivity measurement.\cite{R23,R29,R30} To minimize the error in the half-time method, the time scale of the measurement, $t_{\rm{measure}}$ is required to be sufficiently shorter than $\tau$, typically $\tau$ >> $t_{\rm{measure}}$.

\subsection{\label{sec:level2}Curve fitting method}

When the adiabatic condition is not satisfied, \em{e.g.}\rm, $\tau \sim t_{\rm{measure}}$, one can use the curve fitting method that takes into account the heat leak to the surroundings. Cape and Lehman provided the following equations\cite{R23,R24} to account for the influence of the heat leak on Eq. (1):

\begin{equation}
\begin{split}
T\left(t,L\right)&=T_0+T_{\rm{max}}\left[\frac{2{X_0}^2}{{X_0}^2+2Y+Y^2}\exp{\left(\frac{-{X_0}^2}{\pi^2}\frac{t}{t_0}\right)} \right.\\
&\left.+\sum_{n=1}^{\infty}\left(-1\right)^n\frac{2{X_n}^2}{{X_n}^2+2Y+Y^2}\exp{\left(\frac{-{X_n}^2}{\pi^2}\frac{t}{t_0}\right)}\right],
\end{split}
\end{equation}

\begin{equation}
\begin{split}
X_n \approx n\pi&+\frac{2Y}{\left(n\pi\right)^1}-\frac{4Y^2}{\left(n\pi\right)^3}+\left(\frac{16}{\left(n\pi\right)^5}-\frac{2}{\left[{3\left(n\pi\right)}^3\right]}\right)Y^3 \\
&+\left(\frac{-80}{\left(n\pi\right)^7}+\frac{16}{\left[{3\left(n\pi\right)}^5\right]}\right)Y^4+O\left(Y^5\right),
\end{split}
\end{equation}

\begin{equation}
X_0\approx\left(2Y\right)^{0.5}\left(1-\frac{Y}{12}+\frac{11Y^2}{1440}+O(Y^3)\right),
\end{equation}

\begin{equation}
t_0=\frac{L^2}{\alpha\pi^2},
\end{equation}

where $Y$ is the Biot number expressing the influence of the heat leak into the surroundings. Fig. 3(c) shows the time dependence of ${\Delta}T$/$T_{\rm{max}}$ for several kinds of Biot numbers. If $Y \neq$ 0, ${\Delta}T$/$T_{\rm{max}}$ reaches its peak at $t_{\rm{max}}$ and then decreases exponentially. In this model, $\alpha$ is estimated by fitting Eq. (8) to the experimental data using $\alpha$ and $Y$ as free parameters. For the estimation of $C$, the extrapolated maximum temperature change at $t$ = 0, $T_{\rm{max}}^ *$, is calculated by an exponential fit to the ${\Delta}T(t)$ and its extrapolation at $t$ = 0. Note that the fit should be completed after the peak temperature ($t$ > $t_{\rm{max}}$) where ${\Delta}T$ shows an exponential decay. The heat capacity of the sample can be calculated as below:

\begin{equation}
C=\frac{{\Delta}Q}{T_{\rm{max}}^*} .
\end{equation}

After $\alpha$ is extracted by Eqs. (8)-(11), the $\alpha$ and $C$ can be converted to $\kappa$ and $c_V$ by using Eqs. (4) and (7).

\begin{figure}
\includegraphics{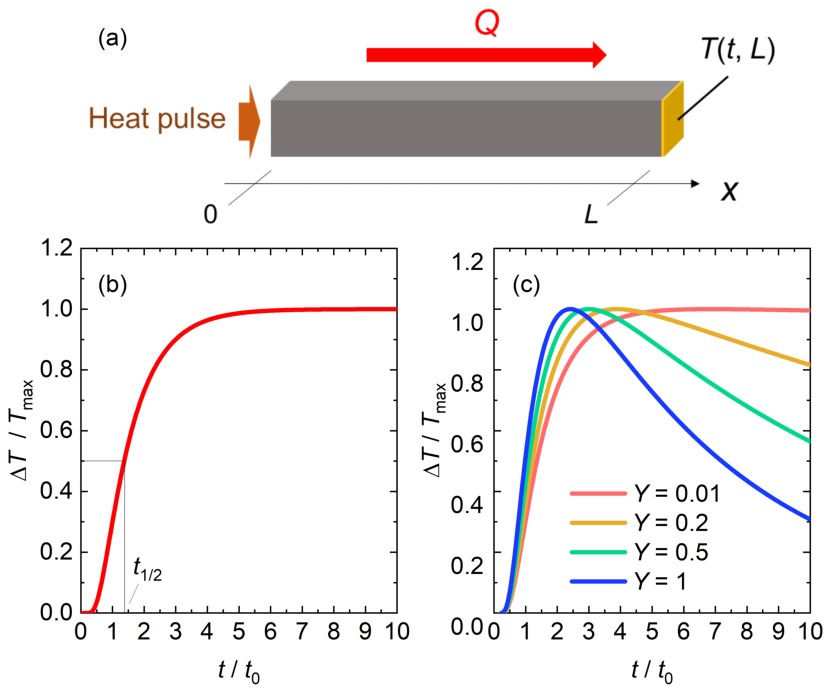}
\caption{\label{fig:epsart1} (a) Simplified one-dimensional heat transfer model. (b) A time profile of temperature rise ${\Delta}T$ calculated by Eq. (1). (c) Time profiles of ${\Delta}T$ calculated by Eq. (8) with different Biot numbers $Y$. The values of ${\Delta}T$ are normalized by the maximum temperature rise $T_{\rm{max}}$.}
\end{figure}

\section{\label{sec:level1}Results and Discussion}
\subsection{\label{sec:level2}Experimental results at zero field}

Figure 4 displays the time profile of the heat-pulse measurement in our setup (setup 1) under zero magnetic field. The red shadowed line denotes the applied power for heating. The starting time ($t$ = 0) is defined as the midpoint of the heat pulse to avoid the heating pulse-duration effect.\cite{R31} After the application of the heat pulse on one end of the sample, the temperature on the backside of the sample rapidly increases from the base temperature, $T_{\rm{0}}$, with the timescale of millisecond and then gradually decreases. From the $T$($t$, $L$) profile, the half time is evaluated as $t_{\rm{1/2}}$ = 5.53 ms, which yields the thermal diffusivity of $\alpha$ = 52.0 mm$^2$s$^{-1}$ using Eq. (6) with $L$ of 1.44 mm. The heat capacity of $C$ = 6.59 JK$^{-1}$mol$^{-1}$ is also obtained using Eq. (12) with $T^*_{\rm{max}}$ of 0.235 K and the applied heat of ${\Delta}Q$ = ${\int}P$($t$)d$t$ = 2.89 {\textmu}J. Then, the thermal conductivity is extracted as $\kappa$ = 4.04 WK$^{-1}$m$^{-1}$ using Eq. (7). As shown as a red curve in Fig. 4, the time profile can be fitted well by Eq. (8) with the fitting parameters of $\alpha$ = 51.8 mm$^2$s$^{-1}$ and $Y$ = 0.0658. The value of $\alpha$ obtained by the curve fitting method agrees well with that of the half-time method, indicating that both methods can be applied in our setup when the adiabatic condition is satisfied. The small $Y$ also confirms that the experiment is close to the adiabatic limit. 

Figure 5(a) shows the temperature dependence of $t_{\rm{1/2}}$ and $\alpha$ in SrCu$_2$(BO$_3$)$_2$ obtained in the setup 1 and 2. Here, setup 2 has the different thermometer sensitivity, resistance of heater, and thermal coupling between the sample and thermal bath, as well as the length of the sample, with those of setup 1. While the $t_{1/2}$ in setup 2 is shorter than that of setup 1 due to the different length of the sample, we confirm that the estimated values of $\alpha$ based on the half-time method (Eq.(6)) show good consistency between the two setups. The thermal diffusivity under zero magnetic field decreases gradually from high temperatures and increases exponentially below 8 K. The increase of $\alpha$ can be attributed to the suppression of spin-phonon scattering due to the spin-gap formation.\cite{R32} We note that the measurable range of $t_{1/2}$ was limited to $\sim$1 ms in this experiment, which restricts the available data region of thermal diffusivity down to 6 K. This is because the duration of the heat pulse, $\tau_{\rm{h}}$, which was chosen to be 0.2 -- 1 ms for this time, should be several times shorter than the $t_{\rm{1/2}}$ ($t_{\rm{1/2}}$ >> $\tau_{\rm{h}}$). This restriction can be removed by measuring a long sample or applying an intense and short pulse current to the heater. Using Eq. (12), the temperature dependence of $C$ is evaluated as seen in Fig. 5(b). The zero-field data show a broadening peak structure around 8 K. The peak structure, which is caused by the spin-gap formation, is quantitatively consistent with the previous study as shown by the blue dashed curve in Fig. 5(b).\cite{R38} Fig. 5(c) shows the temperature dependence of $\kappa$ calculated from the heat capacity and the thermal diffusivity using Eq. (7). The $\kappa$ decreases monotonically from high temperature and turns to increase below 8 K. Such a temperature dependence is widely observed in inorganic spin-gap systems.\cite{R34,R35} The temperature dependence and absolute value of $\kappa$ are also consistent with the reported data as shown by the yellow dashed curve.\cite{R36} These consistencies demonstrate the validity of the present method. Importantly, these measurements were performed within the timescale several times longer than $t_{\rm{1/2}}$ (Fig.5(a)) which is still shorter than the timescale used for a standard steady-state method.

\begin{figure}
\includegraphics{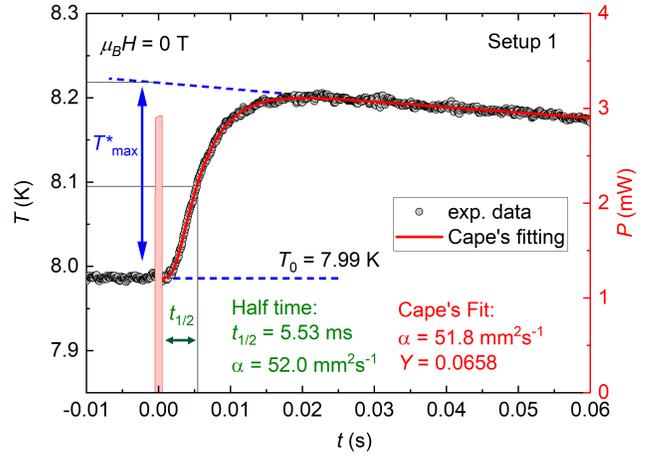}
\caption{\label{fig:epsart1}Example of the time-temperature profile for SrCu$_2$(BO$_3$)$_2$ with the setup 1 under zero magnetic field. The red solid line represents the fitting curve by Eq. (8). After the pulse heating at $t$ = 0, the temperature of the back side rapidly increases from the base temperature, $T_0$, and then gradually decreases exponentially.}
\end{figure}

\begin{figure}
\includegraphics{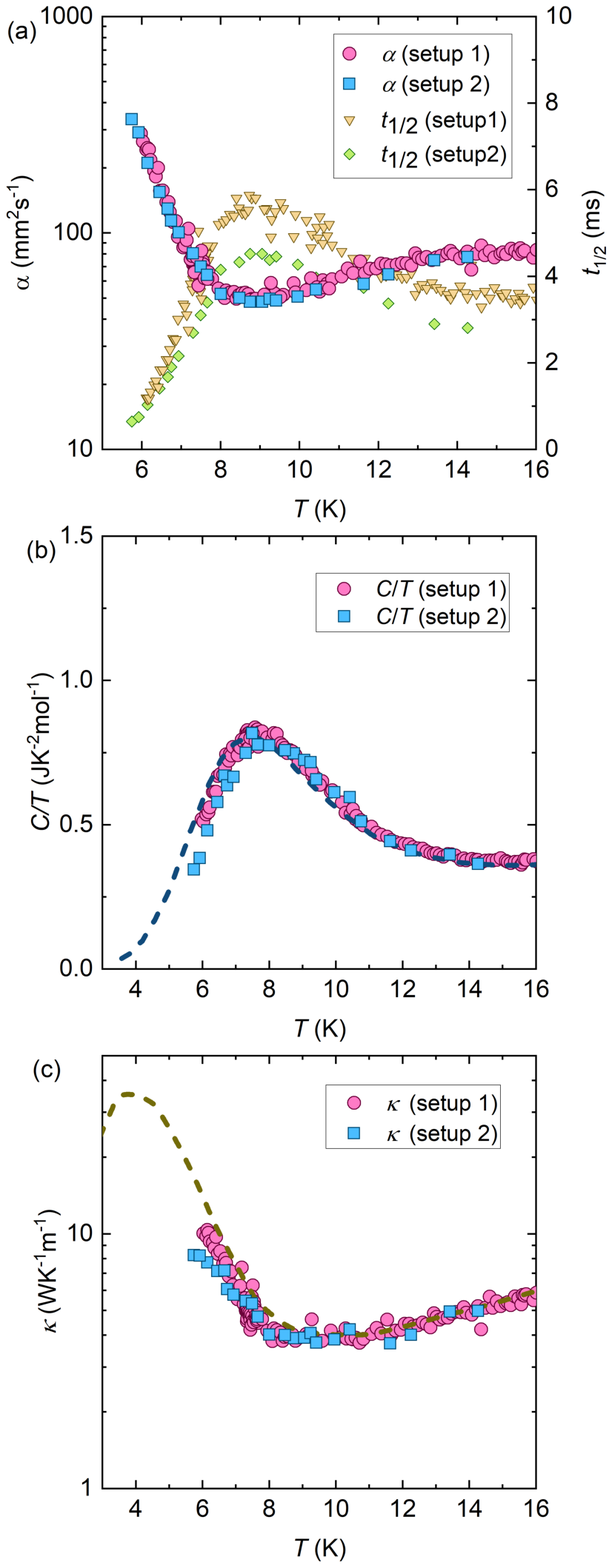}
\caption{\label{fig:epsart1}(a) Temperature dependence of the half-time ($t_{1/2}$) and the thermal diffusivity ($\alpha$) under zero magnetic field. (b) Temperature dependence of the heat capacity under several magnetic fields magnitudes as $C$/$T$ versus $T$ plot. The literature values of SrCu$_2$(BO$_3$)$_2$ from Ref. 36 are also plotted as the blue dashed line. (c) Temperature dependence of the thermal conductivity under zero magnetic field. The literature values of SrCu$_2$(BO$_3$)$_2$ from Ref. 39 are also plotted as the yellow dashed line.}
\end{figure}

\subsection{\label{sec:level2}Experimental results in pulsed magnetic fields}

Figure 6(a) shows the overall time profiles of the sample temperature (brown), magnetic field (blue), and applied power (red shadow) obtained in setup 1. Figure 6(b) shows these enlarged views around the flat-top field region, where the time profile of temperature rise, ${\Delta}T$, with respect to $T_{\rm{0}}$ is plotted. In the measurement shown in Fig. 6(a), a constant power of 0.063 mW was applied before reaching the maximum field to control the measurement temperature. The application of the constant power leads to a temperature rise in the low field region. Then, the rapid increase of magnetic fields at $t$ = -0.45 s induces the drop of the sample temperature up to $\sim$25 T even with the application of constant power, and in turn leads to the gradual increase of temperature above 25 T, which is caused by the strong magnetocaloric effect inherent to SrCu$_2$(BO$_3$)$_2$.\cite{R25} Subsequently, the magnetic field was stabilized to 37.22 T by the mini magnet over the measurement time scale of 50 ms, resulting in the stabilization of the sample temperature, as seen in Figs. 6(a) and (b). It is important to note that the magnetic-field fluctuation in the flattop region is $\pm$0.005 T, which would only cause a small temperature fluctuation of $\pm$0.002 K even in a material with a large magnetocaloric effect of $\sim$0.4 K/T such as a magnetic refrigerant material,\cite{R37} indicating that the field stability is good enough for the present method.

On the flat-top region (Fig.6 (b)), the pulse power was applied at $t$ = 0. The temperature on the backside increases rapidly and shows the local peak around $t$ = 0.03 s, then gradually decreases. The half-time $t_{\rm{1/2}}$ is estimated to be 0.00994 s, yielding $\alpha$ = 29.0 mm$^2$s$^{-1}$ using Eq. (6) with $L$ of 1.44 mm. We also fit the time profile using Eq. (8) shown as the red solid curve in Fig. 6(b), which yields the values of $\alpha$ = 27.9 mm$^2$s$^{-1}$ and $Y$ = 0.182. The fair agreement between both methods suggests that the half-time method also gives a reliable estimation of $\alpha$. Strictly speaking, because of the non-negligible $Y$, the $\alpha$ obtained by the latter fit should be more accurate. The measurement was also performed in setup 2. Fig. 6(c) shows the time dependence of the temperature rise measured in setup 2 at 21.23 $\pm$ 0.005 T. The value of $\alpha$ = 22.0 mm$^2$s$^{-1}$ obtained by the half-time method is again in good agreement with the value of $\alpha$ = 24.1 mm$^2$s$^{-1}$ obtained by the curve fitting method, which suggests that ideal measurement conditions are satisfied regardless of the magnetic field, temperature, and sample size. From the data shown in Fig. 6(b) and (c), the heat capacity under magnetic fields can be calculated. For example, the analysis of the data shown in Fig. 6(b) yields $C$ = 1.48 and 1.30 JK$^{-1}$mol$^{-1}$ using Eqs. (3) and (12), respectively. These values allow us to evaluate the absolute values of the thermal conductivity of $\kappa$ = 0.513 and 0.445 WK$^{-1}$m$^{-1}$ at 3.26 K and 37.2 T using Eq. (7). Although there is an ambiguity originating from the choice of the analysis methods, thermal diffusivity, heat capacity, and thermal conductivity can be obtained from single time-temperature curve taken in pulsed magnetic fields. Since the heat capacity obtained using Eq. (3) does not include the effect of heat losses to the outside, the value of thermal conductivity obtained by the curve fitting method is considered to be more accurate.

\begin{figure}
\includegraphics{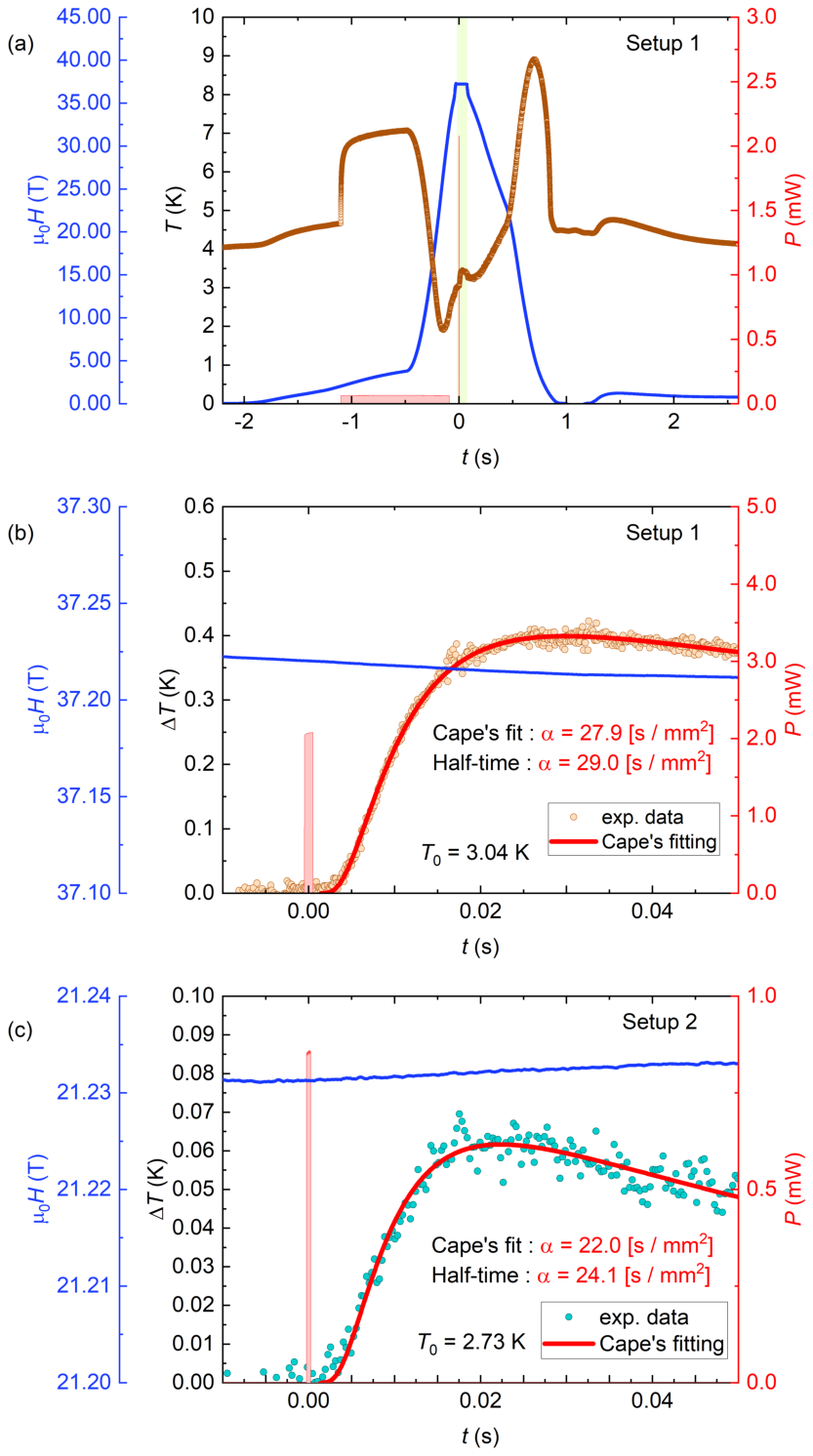}
\caption{\label{fig:epsart1}(a) Example of the time-temperature profile for SrCu$_2$(BO$_3$)$_2$ with setup 1. The flat-top field was 37.2 T and maintained for $\sim$50 ms (yellow region). The blue solid curve and the red shadowed area denote the time dependence of the applied magnetic fields and the applied power, respectively. (b) The enlarged figure of the flat-top field region in (a). The red solid line is the fitting curve by Eq. (8). (c) Another example of the time profile with setup 2.}
\end{figure}

In Fig. 7, the analyzed temperature dependences of $\alpha$, $C$/$T$, and $\kappa$ taken in setup 1 (circles) and setup 2 (squares) are shown together with previous reports (dashed curve).\cite{R33, R36, R39, R38} The reduction of $\alpha$ that is equivalent for the long $t_{\rm{1/2}}$ is observed above 24 T, while $\alpha$ shows a strong increase below 12 T. Because of the timescales of the flat-top region and of the heat pulse, the measurable timescale of $t_{\rm{1/2}}$ is roughly from 0.2 to 20 ms in this experiment, which restricts the measurable temperature range of $\alpha$ down to 2 K; $\alpha$ = 1000 mm$^2$s$^{-1}$ for the setup 1 and setup 2 corresponds to $t_{\rm{1/2}}$ = 0.288 and 0.217 ms, respectively, and $\alpha$ = 20 mm$^2$s$^{-1}$ for the setup 1 and setup 2 corresponds to 14.4 and 10.8 ms, respectively. The low-temperature shift of the broad peak and the upturn structure are observed in $C$/$T$ with increasing magnetic fields (FIg.7(b)). These field dependences in $C$/$T$ are likely caused by the suppression of the spin gap and the subsequent emergence of the first-order phase transition. In the present case, the increase of $C$ with dropping temperature might be caused by the tail structure of the field-induced first-order phase transition located outside our measurement window.\cite{R40, R25} Except for the high-temperature structure at 24.8 T, the $C$/$T$ curves in zero and finite magnetic fields are in good agreement with the reported data,\cite{R38, R39} both in absolute value and in the relative temperature dependence. The deviation of the high-temperature data at 24.8 T from the reported data at 24 T is probably due to a small sample misalignment, accumulated errors in the temperature calibration, or intrinsic field dependence in $C$/$T$. A peak structure observed in $\kappa$ at 0 T is suppressed by the application of magnetic fields (Fig. 7(c)). The field and temperature dependences in $\kappa$ also show fair agreements with  earlier reports.\cite{R33} The broad zero field peak in $\kappa$ and its shift to a lower field are caused by the emergence and its suppression of the spin gap.\cite{R33} The suppression of the spin gap is expected to reduce the mean free path of phonons and suppress the zero field peak structure in $\kappa$. The weak field dependence in $\kappa$ above 20 T is similar to another spin gap system, BiCu$_2$PO$_6$,\cite{R41} where $\kappa$ is dramatically suppressed below the critical field of the spin gap closure and becomes nearly constant above the critical field. This scenario should be also realized in SrCu$_2$(BO$_3$)$_2$ and is consistent with the observed weak field-dependence of $\kappa$. These agreements demonstrate that the present method can indeed be used to investigate thermal conductivity in pulsed high magnetic fields.

\begin{figure}
\includegraphics{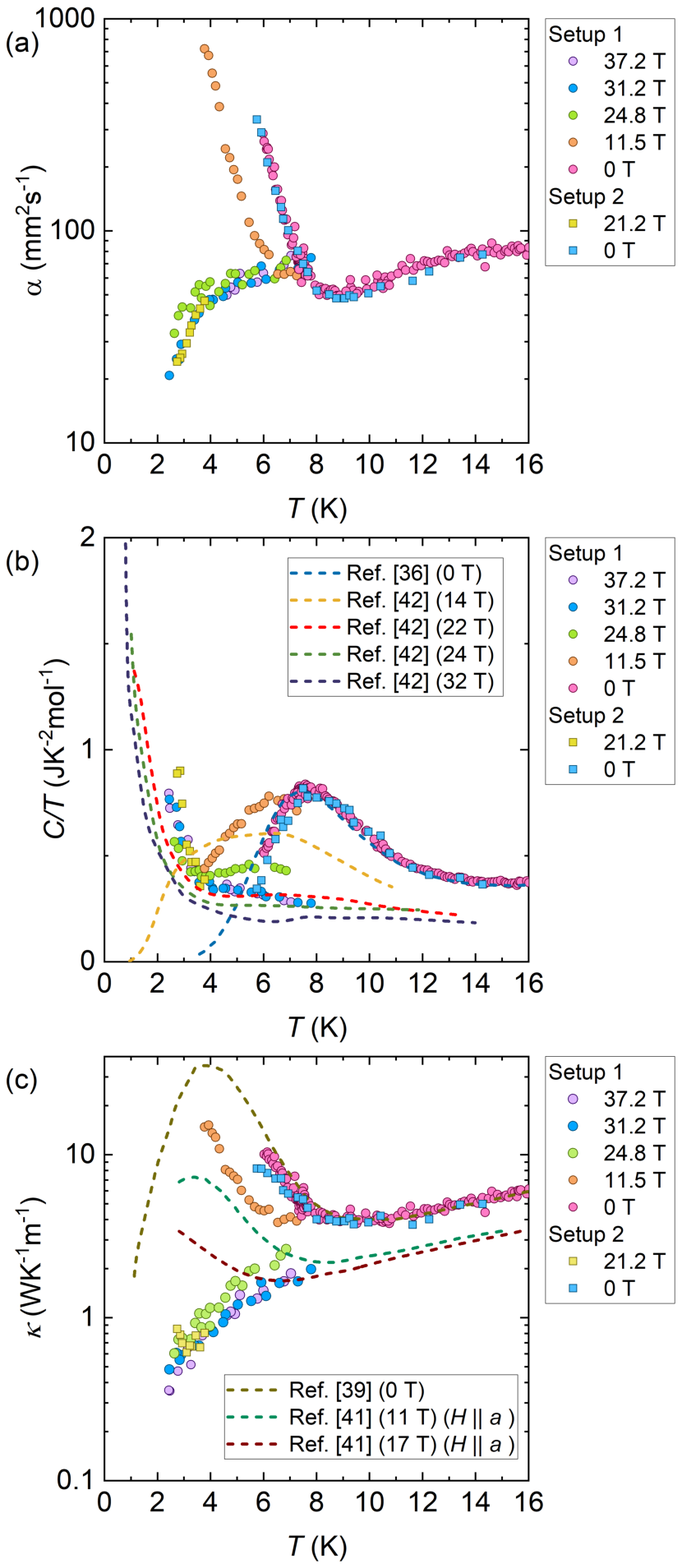}
\caption{\label{fig:epsart1}(a) Temperature dependence of the thermal diffusivity under different magnetic fields. (b) Temperature dependence of the heat capacity under different magnetic fields as $C$/$T$ versus $T$ plot. (c) Temperature dependence of the thermal conductivity under different magnetic fields. The literature values of SrCu$_2$(BO$_3$)$_2$ are also plotted as dashed curves.}
\end{figure}

In summary, we demonstrate that the flash method can be used for the simultaneous measurement of specific heat and thermal conductivity in highly stabilized pulsed magnetic fields of up to $\sim$37.2 T. There are many electronic and magnetic phases leading to anomalous heat conduction in low magnetic fields, such as magnon Bose-Einstein condensation (BEC) states\cite{R42,R43} and charge-neutral fermions in Kondo insulators.\cite{R5} By extending the thermal conductivity measurement of these materials to higher magnetic fields using the present technique with long pulse magnets,\cite{AR3,AR4,AR5} it will be possible to study novel and anomalous heat transport phenomena in more detail.

\section{\label{sec:level1}Conclusion}

We performed simultaneous measurements of thermal diffusivity, heat capacity, and thermal conductivity based on the flash method under pulsed high magnetic fields  up to $\sim$37 T. We have obtained the thermal properties on the test single crystal sample of SrCu$_2$(BO$_3$)$_2$. The temperature changes in the sample after the pulse heating can be fitted to the theoretical curve proposed by Cape and Lehman, and the extracted values of thermal properties are in agreement with previous reports, which supports the validity of the present method. Our method is a strong tool for comprehensive studies of heat transport and thermodynamic phenomena in pulsed magnetic fields as well as in static fields owing to the convenience of fast measurement time.

\section{\label{sec:level1}ACKNOWLEDGMENTS}

We thank Dr. Albert Migliori, Los Alamos National Laboratory, for his encouragement and suggestions. We also thank Dr. Kazuki Matsui, Institute for Solid State Physics, The University of Tokyo, for support for the equipment used in our experiments. This work was partly supported by the Precise Measurement Technology Promotion Foundation (PMTP-F), JSPS KAKENHI Grant Number 22H00104, and 21K20347. Work at the pulsed field facility of the National High Magnetic Field Laboratory was supported by the National Science Foundation Cooperative agreements DMR-1644779 and DMR-2128556, the State of Florida, and the US DOE Basic Energy Science project "Science at 100T".

\section{\label{sec:level1}Data availability}

The experimental data, circuit design, and source codes that support the findings of this study are available from the corresponding author upon reasonable request.

\nocite{*}
\bibliography{aipsamp}

\end{document}